\documentstyle[12pt,draft,epsf]{article}
\newcommand{\beq}{\begin{equation}}
\newcommand{\eeq}{\end{equation}}

\newcommand{\unmezzo} {\frac{1}{2}}
\newcommand{\unquarto}{\frac{1}{4}}
\newcommand{\unottavo}{\frac{1}{8}}

\newcommand{\tra}{ \mbox{ Tr~} }
\newcommand{\erf}{ \mbox{ erf} }

\begin{document}
\pagenumbering{arabic}

\title{ LATTICE GAS ANALOGUE OF SK MODEL:\\
A paradigm for the glass transition }

\author{ Francesco M. RUSSO
\\[1.0em]
Dipartimento di Fisica,Universit\`a di Roma {\it Tor Vergata},\\
Via della Ricerca Scientifica 1, 00133 Roma, Italy\\
and\\ INFN, Sezione di Roma }

\date{ February 10, 1998}

\maketitle

\begin{abstract}

We investigate the connection between the well known Sherrington-Kirkpatrick
Ising Spin Glass and the corresponding Lattice Gas model by analyzing the
relation between their thermodynamical functions. We present results of replica
approach in the Replica Symmetric approximation and discuss its stability as a
function of temperature and external source. Next we examine the effects
of first order Replica Symmetry Breaking at zero temperature. We finally
compare SK results with ours and suggest how the latter could be relevant
to a description of the structural glass transition.

\end{abstract}

\vfill
{\bf \hfill ROM2F/97/44 }\\
\hfill To appear in ~{\em Journal of Physics\/}{\bf A}

\newpage
\section{Introduction}

 ~ It is well known\cite{huang} that Ising Model is equivalent to Lattice Gas,
a system defined in terms of occupation variables $\tau$ taking values $0$ and
$1$. The Lattice Gas effective Hamiltonian is formally identical to Ising
one\cite{huang}. A simple change of variables ($\sigma=2\tau-1$) maps each of
the two Hamiltonians into the other, provided that Ising external field is
related to lattice gas chemical potential by $h-J=\unmezzo\mu$, where $J$ is
Ising spin-coupling related to lattice gas site-coupling $\Phi$ by
$J=\unquarto\Phi$. This results in a simple relation between Ising free
energy density and Lattice Gas pressure: $p=h-\unmezzo J-f$. The two systems 
have therefore the same phase diagram and the same critical behaviour (real
gases and Ising magnets are in the same universality class).

 For random systems\cite{MPV} this whole argument breaks down because the
relation between chemical potential and magnetic field involves the quenched
couplings. As we shall see in the next sections this results in new and
unexpected features for the phase diagram of the system.

  For Neural Networks this inequivalence between spin($\pm 1$) and 
occupation ($0$,$1$) variables had already been pointed out and analyzed,
see for example \cite{tsofei} and references therein.

Recently much effort has been devoted to develop a description of the
structural glass transition\cite{BOUMEZ,MAPARI1,MAPARI2} within the framework
of disordered systems. Much of these models were however based on Ising Spin
variables instead of Lattice Gas ones which should be more appropriate for
a condensed matter system. For disordered systems the two kinds of variables
are not equivalent. To have a comparison term it would be useful to analyze
the properties of a mean field disordered lattice gas model. This could also
show us how the structural glass transition may be described in the same
way as the liquid-gas one.

 In section $2$ we present our model and in the following one we analyze its
ground state. In section $4$ we discuss its relation with the SK model. Section
$5$ is dedicated to writing down the saddle point equations of the replica
approach. In section $6$ we analyze the low temperature behaviour of the
Replica Symmetric approximation and show its {\em Phase Diagram\/}.
A condensed version of section $5$ and $6$ was originally presented in
ref. \cite{RUSSO0}.   In section
$7$ we discuss the stability of the Replica Symmetry and look for the AT line
of the model. Section $8$ contains some preliminary results with broken Replica
Symmetry and in section $9$ we present a comparison of our main results with
the SK's ones and draw some tentative conclusions.

\section{The Model}

~  We consider a system of $N$ sites. An occupation variable $\tau_k$, defined
in each site $k$, can take the values $0$ or $1$. The Hamiltonian of
the system is taken to be formally identical to the SK's one\cite{SK1,SK2}. The
interaction energy between two different ($k$ and $l$) occupied sites is taken
to be $\phi_{kl}$ and the system is coupled to some external source $g$. The
total effective Hamiltonian is therefore:
\beq
 H_\phi[\tau]=-g\sum_{k=1}^{N}\tau_k-\unmezzo\sum_{l\neq k}\phi_{kl}\tau_k\tau_l
   ~.
\label{Ham}
\eeq

In magnetic language $g$ would be the external field, while for a Lattice Gas
$g$ is the sum of the chemical potential, the kinetic contribution, and
eventually an external force term.

The infinite-ranged interaction energies $\{\phi_{kl}\}$ are taken to be
quenched independent Gaussian variables with zero mean and variance $\Phi^2/N$.
This means that the model should not have, on average, a preference between an
ordering transition in two sublattices, as in binary alloys, and gas-liquid
transition. For the low-temperature behaviour we could expect al least two
regimes. For a strong enough negative $g$ occupied sites should be energetically
penalized and we expect, for most of the realizations of the noise, the ground
state energy to be zero and the low temperature behaviour to be similar
to that of a pure system. On the other hand, for a strong enough positive $g$
occupied sites should be energetically favoured and we could expect the low
temperature behaviour of the system to be much similar to that of SK's one.
In the following we shall take $\Phi$ as the unit of energy and set
$\Phi\equiv 1$, so we have:
\beq
 P(\phi_{lk})=\left(\frac{N}{2\pi}\right)^\unmezzo
 \exp\left[-\unmezzo N\phi_{lk}^2\right].
\label{P(phi)}
\eeq
Each $\phi_{kl}$ is taken to be equally distributed and
therefore each site interacts with each other.
We also set the Boltzmann constant equal to $1$, as
a consequence $T,H,g$ and $\phi$ are all dimensionless.

For a given realization of the $\phi$'s the partition function is:
\beq
 Z_\phi(\beta;g)=\sum_{\{\tau\}}e^{-\beta H\phi[\tau]}~.
\label{Zphi}
\eeq

For a Lattice Gas this would be the Grand Canonical partition function, because
the total number of occupied sites $\sum_k\tau_k$ is not constrained to a specific
value. Strictly speaking we should call ``pressure'' the thermodynamic potential
($\ln Z$), while the grandcanonical independent variable should be the chemical
potential. To have a more immediate comparison with SK model, and also with
other models based on spin variables, we shall however abusively refer to the
thermodynamical variables as if they were the canonical ones. In the following
we shall therefore call ``free energy density'' the thermodynamic potential
and refer to $g$ as the ``external field''.

We are interested, as usual when dealing with quenched disorder,
to evaluate the averaged free energy density:
\beq
 f(\beta;g)=-\frac{1}{\beta N}\int P[\phi]\ln Z_\phi d\phi ~,
\label{f-med}
\eeq
this will be done in the following using the replica approach\cite{MPV,EA,SK1}.

\section{ Ground State Properties }
 
 We now discuss in more detail the ground state picture which was conjectured
in the previous section. First of all we observe that the energy of the
configuration without occupied sites is zero for all realizations of the
quenched couplings. We can therefore conclude that the ground state energy is
always non-negative. Secondly, we can easily see that the configuration with
all occupied sites has an average energy equal to $-gN$, which is exactly the
same value obtained for the all-spins-up configuration of SK model. Combining
these two results we obtain, for the (quenched averaged) ground state energy
density, the following upper bound:
\[
  u \leq -g\theta(g)
\]

 A lower bound can be obtained from the knowledge of the eigenvalues for a
large Gaussian random matrix. Let us call $\omega$ the maximum eigenvalue
of the matrix $[\phi_{lk}]$. For every configuration we can write
\[
   \sum_{lk} \tau_l\phi_{lk}\tau_k \leq \omega \sum_k (\tau_k)^2 ~.
\]
In our case we have $(\tau_k)^2=\tau_k$, and $\omega=2$ with probability one
in the thermodynamic limit. Putting all together we obtain, for the
Hamiltonian (\ref{Ham}), the following bound:
\[
  -(1+g)\sum_k \tau_k \leq  H_\phi[\tau]  ~.
\]
 Taking the minimum of both sides we finally get:
\[
  -(1+g)\theta(g+1) \leq  u  ~.
\]

We thus see that for $g\leq -1$ the upper and lower bounds saturate and
$u=0$ with probability one in the thermodynamic limit. The ground state
configuration is that with no occupied sites. Each configuration with a finite
number of occupied sites has the same average energy density in the
thermodynamic limit. We therefore conclude that the low temperature
behaviour of the system should be very similar to that of a pure one.

On the other hand, for positive $g$ we have $u\leq -g$, bound which is
saturated in the limit $g\to\infty$. This implies that our system has a
(first order?) zero-temperature transition at some $g_0$ between $-1$ and $0$.
We also speculate that for a large positive $g$ the low temperature behaviour
of the system should approximate that of SK model. We note that in SK case
repeating the previous arguments we would obtain $-1-|h|\leq u\leq -|h|$.

The lower limit for $g_0$ can be improved using Derrida's argument\cite{REM}.
Let us define $\Sigma(g,N)$ the average number of configurations with negative
energy for a system with $N$ sites:
\[
  \Sigma(g,N) = \int P[\phi]\sum_{\tau}\theta(-H_\phi[\tau])d\phi ~.
\]
For a negative $g$ we obtain, in the thermodynamic limit,
\[
  \sigma \equiv\lim_{N\to\infty} \frac{1}{N}\ln\Sigma(g,N)
   = \ln 2 - g^2  ~.
\]
If $g< -\sqrt{\ln 2}$ we see that $\sigma<0$ and therefore $\Sigma(g,N)\to 0$
in the thermodynamic limit. Following Derrida\cite{REM} we can conclude that
for $g< -\sqrt{\ln 2}=-0.8326$ there are almost surely no configurations of
negative energy in the thermodynamic limit. Since there always is at least a
configuration of zero energy, this must be ground state energy. We thus obtain
the lower bound $g_0\geq -\sqrt{\ln 2}=-0.8326$ for the transition field at
zero temperature.

\section{Relation with SK model}

 ~ Before proceeding in our analysis we want to discuss the connections between
the SK model\cite{SK1} and the previously introduced one.  We assume, as for
the corresponding homogeneous models\cite{huang}, that site variables are
related by:
\beq
 \sigma=2\tau-1 ~~~~~~\Longleftrightarrow ~~~~~ \tau=\unmezzo(\sigma+1) ~.
\label{cambio}
\eeq
Substituting (\ref{cambio}) into the Hamiltonian (\ref{Ham}) we get:
\beq
 -H_\phi\left[\frac{\sigma+1}{2}\right]=
 \unmezzo\sum_k\left(g+\unquarto\Phi_k\right)
 +\sum_k\left(\unmezzo g+\unquarto\Phi_k\right)\sigma_k
 +\unottavo\sum_{k\neq l}\phi_{kl}\sigma_l\sigma_k ~,
\label{Hditdis}
\eeq
where:
\[
 \Phi_k=\sum_l^{l\neq k}\phi_{kl} ~.
\]
We stress that (\ref{Hditdis}) connects a homogeneous (site-independent) field
system to a local-field one. In fact, defining the SK effective Hamiltonian
for a site-dependent magnetic field:
\beq
 H_{J,h}^{(SK)}[\sigma]=-\sum_k h_k\sigma_k-
 \unmezzo\sum_{k\neq l}J_{kl}\sigma_l\sigma_k ~,
\label{H-SK}
\eeq
we can write:
\beq
 H_\phi\left[\frac{\sigma+1}{2}\right]=
 -\unmezzo\sum\left(g+\unquarto\Phi_k\right)
 +\unquarto H_{\phi,h[\phi]}^{(SK)}[\sigma] ~,
\label{HdiH}
\eeq
and the local magnetic field is correlated to the couplings by:
\beq
 h_k[\phi]=2g+\Phi_k ~.
\label{hdiphi}
\eeq
The change of variables (\ref{cambio}) maps our Hamiltonian (\ref{Ham})
into an SK-like one, but correlation given by (\ref{hdiphi}) is enough to destroy
their equivalence. In fact, summing over configuration, we get for the partition
functions:
\beq
 Z_\phi(\beta;g)=
 \exp\left[\unmezzo\beta\sum_k\left(g+\unquarto\Phi_k\right)\right]
 Z^{(SK)}_\phi\left(\frac{\beta}{4};h[\phi]\right]
\label{ZdiZ}
\eeq
and thus the relation between the free energy densities of the two models leads
to:
\beq
 f(\beta;g)=-\unmezzo g+
 \unquarto\int f^{(SK)}_\phi\left(\frac{\beta}{4};h[\phi]\right]P[\phi]d\phi ~.
\label{fdif}
\eeq
Our system is therefore equivalent to an SK-like one in which the magnetic field
is a local random variable correlated with the couplings. SK's averaged free
energy is not directly related to ours.
Relation (\ref{cambio}) is thus not useful in investigating thermodynamics of
our model that is not reducible to something known, we have to face it by
itself.

\section{The Replica Symmetric Solution}

~ Let us now proceed in applying replica formalism\cite{MPV} to our system; we
have to calculate the averaged $n$-th power of the partition function:
\beq
 Z_n=\overline{(Z_\phi)^n}=\int(Z_\phi)^nP[\phi]d\phi ~.
\label{Zn-def}
\eeq
 For integer $n$ we get, after performing Gaussian integration:
\[
 Z_n=\sum_{\{\tau\}}\exp{\left[\sum_k\beta g\sum_a\tau_k^a+
 \frac{\beta^2}{2N}\sum_{k<l}\left(\sum_a\tau_l^a\tau_k^a\right)^2\right]}~,
\]
using the identity:
\beq
 2\sum_{k<l}\left(\sum_a\tau_l^a\tau_k^a\right)^2=
 \sum_{a,b}\left(\sum_k\tau_k^a\tau_k^b\right)^2-
 \sum_k\sum_{a,b}\tau_k^a\tau_k^b
\label{identau}
\eeq
which follows from the relation $\tau^2=\tau$, we can reorder the exponent and
obtain:
\beq
 Z_n=\sum_{\{\tau\}}\exp\left[\beta\sum_k\left(g\sum_a\tau_k^a
 -\frac{\beta}{4N}\sum_{a,b}\tau_k^a\tau_k^b\right)\right]
 \prod_{a,b}\exp\left[\frac{4}{N\beta^2}
 \left(\frac{\beta^2}{4}\sum_k\tau_k^a\tau_k^b\right)^2\right]~.
\label{zeta-n1}
\eeq
Using Gaussian identities we rewrite (\ref{zeta-n1}) as:
\[
 Z_n=\sum_{\{\tau\}}\exp\left[\beta\sum_k\left(g\sum_a\tau_k^a
 -\frac{\beta}{4N}\sum_{a,b}\tau_k^a\tau_k^b\right)\right]
 \times
\]
\beq
 \times
 \prod_{a,b}\int\left(\frac{N\beta^2}{4\pi}\right)^\unmezzo
 \exp\left(-\frac{N}{4}\beta^2 Q_{ab}^2+
 \unmezzo\beta^2\sum_k Q_{ab}\tau_k^a\tau_k^b\right)dQ_{ab}~,
\label{zeta-n2}
\eeq
reordering the exponentials and defining:
\beq
 H_Q[\tau]=-\unmezzo\beta^2\sum_{a,b}
 \left(Q_{a,b}-\frac{1}{2N}\right)\tau^a\tau^b
 -\beta g\sum_a\tau^a ~,
\label{HQ[tau]}
\eeq
\beq
 A[Q]=\frac{\beta^2}{4}\sum_{a,b}Q_{ab}^2 ~
 -\ln\left[\sum_{\{\tau\}}e^{-H_Q[\tau]}\right]~,
\label{A[Q]}
\eeq
we finally get:
\beq
 Z_n(\beta;g)=\left(\frac{N\beta^2}{4\pi}\right)^\frac{n^2}{2}
 \int e^{-NA[Q]}~d^{n^2}Q ~.
\label{zeta-n3}
\eeq
 The averaged free energy density is given by:
\beq
 f(\beta;g)=\lim_{N\to\infty}\lim_{n\to0}
 -\frac{1}{\beta nN}\ln Z_n(\beta;g)~.
\label{free-lim}
\eeq
In the thermodynamic limit the integral can be estimated by maximizing the
integrand, and this yields:
\beq
 f_n(\beta;g)=\lim_{N\to\infty}-\frac{1}{\beta nN}\ln Z_n(\beta;g)
 =\frac{1}{\beta n}\inf_Q\{A[Q]\}~.
\label{effe-n}
\eeq
The extremum is determined from the saddle point equation:
\beq
 \frac{\partial A}{\partial Q_{ab}}=
 \frac{\beta^2}{2}Q_{ab}-\frac{\beta^2}{2}
 \frac{\sum_\tau\tau^a\tau^b e^{-H_Q[\tau]}}
 {\sum_\tau e^{-H_Q[\tau]}}=0 ~,
\label{Qsp}
\eeq
that may be rewritten as $Q_{ab}=\langle\tau^a\tau^b\rangle_Q$.

 We first consider saddle points that are symmetric under the Replica
Group\cite{MPV}. Setting $Q_{ab}=q+b\delta_{ab}$ we can write
$\sum_{ab}Q_{ab}\tau^a\tau^b=q\left(\sum_a\tau^a\right)^2+b\sum_a\tau_a$ and
$\sum_{ab}Q_{ab}^2=n(q+b)^2+n(n-1)q^2$, substitution in (\ref{A[Q]}) and
extraction of the $n\to0$ limit then yields:
\beq
 f=\unquarto\beta b(2q+b)-(2\pi\beta^2)^{-\unmezzo}\int_{-\infty}^{\infty}
 \ln\left[1+e^{\beta(\alpha+z\sqrt{q})}\right]e^{-\unmezzo z^2}~dz~.
\label{f-sim}
\eeq
In equation (\ref{f-sim}) we set $\alpha=g+\unmezzo\beta b$, and the matrix
elements satisfy the coupled equations:
\[
 \rho\equiv q+b=(2\pi)^{-\unmezzo}\int
 \left[1+e^{\beta(\alpha+z\sqrt{q})}\right]^{-1}e^{-\unmezzo z^2}~dz
\]
\beq
 q=(2\pi)^{-\unmezzo}\int
 \left[1+e^{\beta(\alpha+z\sqrt{q})}\right]^{-2}e^{-\unmezzo z^2}~dz ~.
\label{Qsp-sim}
\eeq

As can be seen following the line of \cite{SK1,SK2} the physical
significance of $\rho$ and $q$ is:
\beq
 \rho=\overline{\langle\tau\rangle}~~~~~~~~~~
 q=\overline{\langle\tau\rangle^2}~,
\label{sign-sim}
\eeq
where, following the notations of \cite{MPV}, a bar denotes
the average over quenched disorder.

 For $\beta=0$ we have $\rho=\unmezzo$ and $q=\unquarto$, as we expect from
their physical significance. In the high temperature regime we can solve
(\ref{Qsp-sim}) by expansion in powers of $\beta$, and this yields:
\[
 \rho=\unmezzo+\unquarto\beta g+\frac{1}{32}\beta^2
\]
\[
 q=\unquarto+\unquarto\beta g+
 \frac{1}{16}\left(\frac{3}{4}+g^2\right)\beta^2 ~.
\]

\section{Low temperature results}

~ At zero temperature we can perform a detailed analytic study of saddle point
equations. In this limit eq. (\ref{Qsp-sim}) leads to:
\beq
 q=(2\pi)^{-\unmezzo}\int_{-\infty}^{\alpha/\sqrt{q}}e^{-\unmezzo z^2}~dz
 \equiv\erf\left(\frac{\alpha}{\sqrt{q}}\right)
\label{Qsp-T0}
\eeq
\[
 \gamma_0\equiv\lim_{T\to0}\beta(\rho-q)=
 (2\pi q)^{-\unmezzo}~e^{-\frac{\alpha^2}{2q}}~,
\]
the equation for $\rho$ is the same as that for $q$ indeed for $T\to0$ we have
$\rho=q+\gamma T$ and thus $\alpha=g+\unmezzo\gamma_0$.\\
For $g<0$ there always is a solution of (\ref{Qsp-T0}) with $q=0$,
$\gamma_0=0$ and $\alpha\equiv g$. 
\begin{figure}
 \epsfxsize=400pt\epsffile{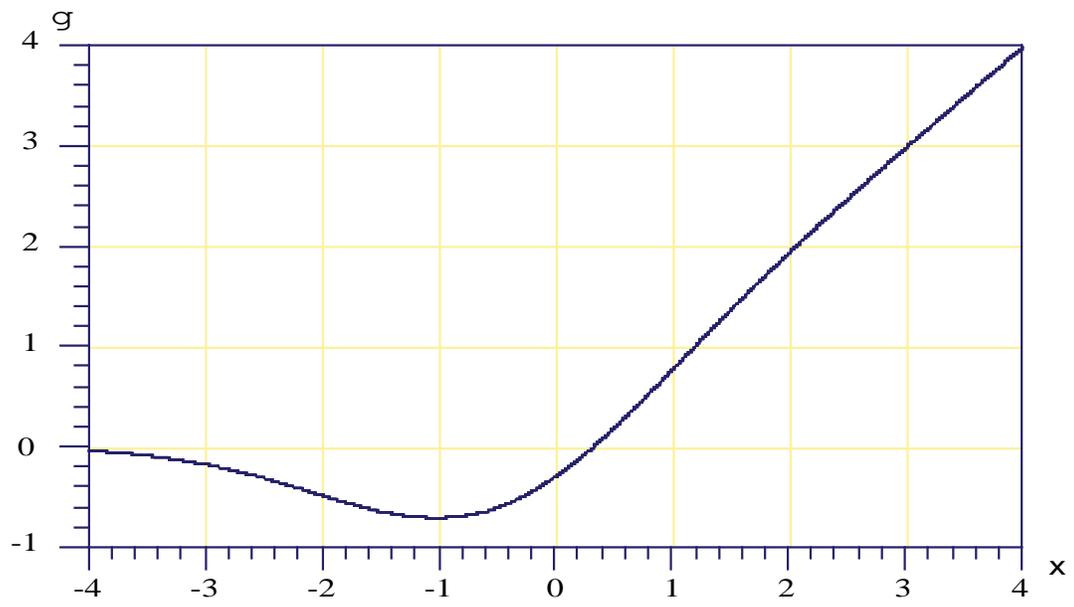}
\caption{External field $g$ versus parameter $x$ ($x\equiv\erf^{-1}(q)$) as
results from the last of (\ref{Qsp-solT0}).}
\label{g-vs-x}
\end{figure}
To look for solutions with $q\ne0$ let us define
$x\equiv\erf^{-1}(q)=\alpha/\sqrt{q}$, then we get:
\[
 \rho=q=\erf(x)
\]
\[
 \gamma_0=\left[2\pi\erf(x)\right]^{-\unmezzo}e^{-\unmezzo x^2}
\]
\[
 \alpha=x\sqrt{\erf(x)}
\]
\beq
 g=\alpha-\unmezzo\gamma_0=
 x\sqrt{\erf(x)}-\unmezzo\left[2\pi\erf(x)\right]^{-\unmezzo}e^{-\unmezzo x^2}~,
\label{Qsp-solT0}
\eeq
we can therefore express all relevant quantities as functions of parameter $x$.
From the last of (\ref{Qsp-solT0})'s (see fig. \ref{g-vs-x}) we can see that
each value of $g$ in the range $-0.70242<g<0$ corresponds to two values of
$x$ both lower than $0.30859$.
\begin{figure}
 \epsfxsize=400pt\epsffile{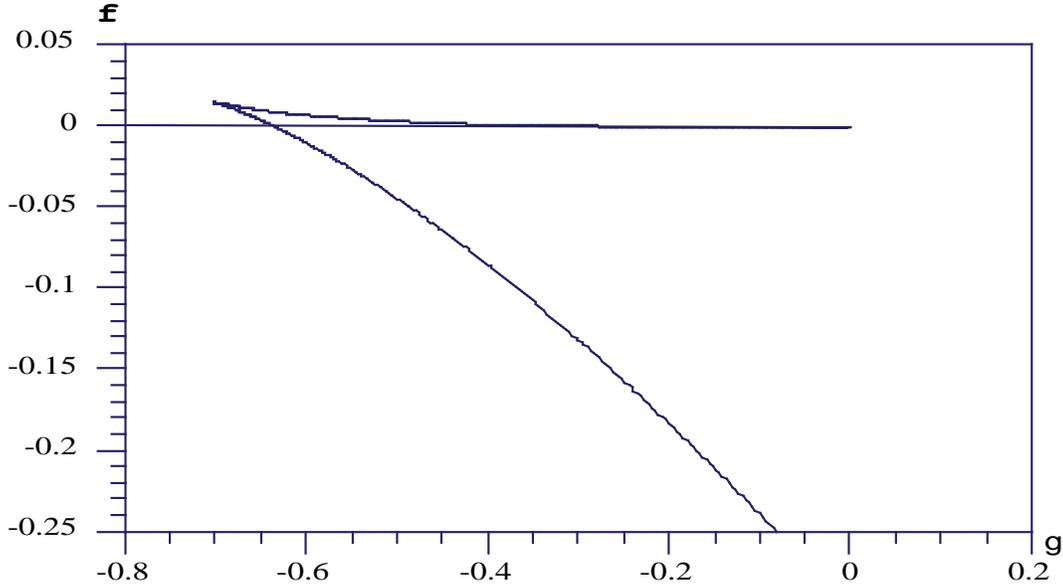}
\caption{Free energy density versus external field at zero temperature in
the replica symmetric solution. The transition point is where the two lower 
branches cross each-other. The higher branch is an unstable solution.}
\label{f0-vs-g}
\end{figure}
Summarizing, if $g<-0.70242$ or $g>0$ we have a single solution of
(\ref{Qsp-T0}) for each value of $g$, but if $-0.70242<g<0$ we have
three solutions and in order to pick-up the physical one we have to
impose the continuity of the free energy density as a function of
external field, see fig. \ref{f0-vs-g}.
\begin{figure}
 \epsfxsize=400pt\epsffile{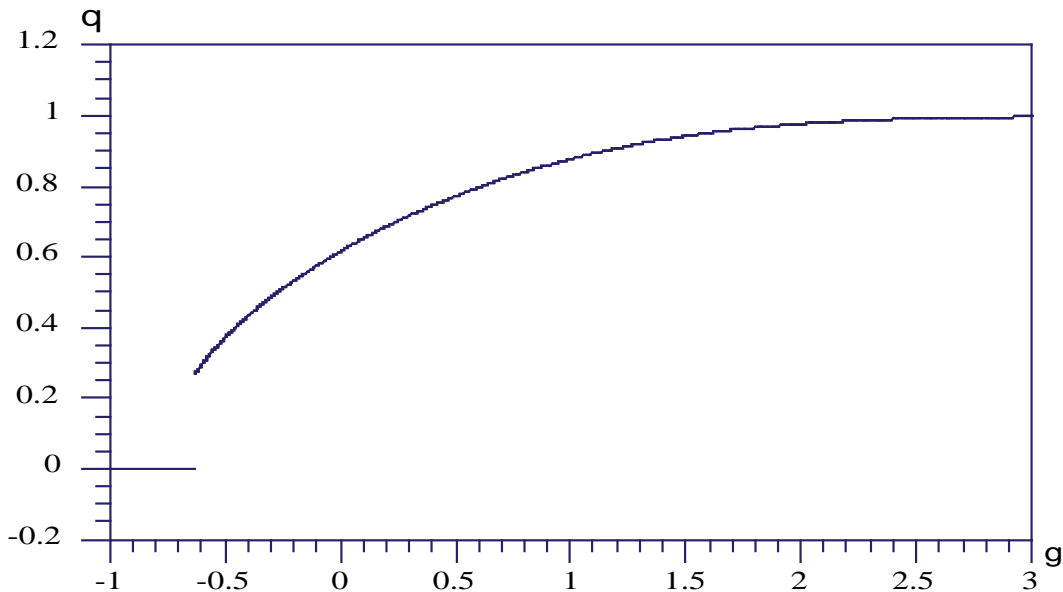}
\caption{Replica symmetric order parameter $q$
($=\rho=\overline{\langle\tau\rangle}$~) at $T=0$
as a function of external field.}
\label{q0-vs-g}
\end{figure}
We thus find a \em first-order phase transition \em in the point $g_0=-0.63633$
where the two lower branches of $f$ cross each-other, the transition point
being determined by the condition $\gamma_0(g_0)=-g_0$.
\begin{figure}
 \epsfxsize=400pt\epsffile{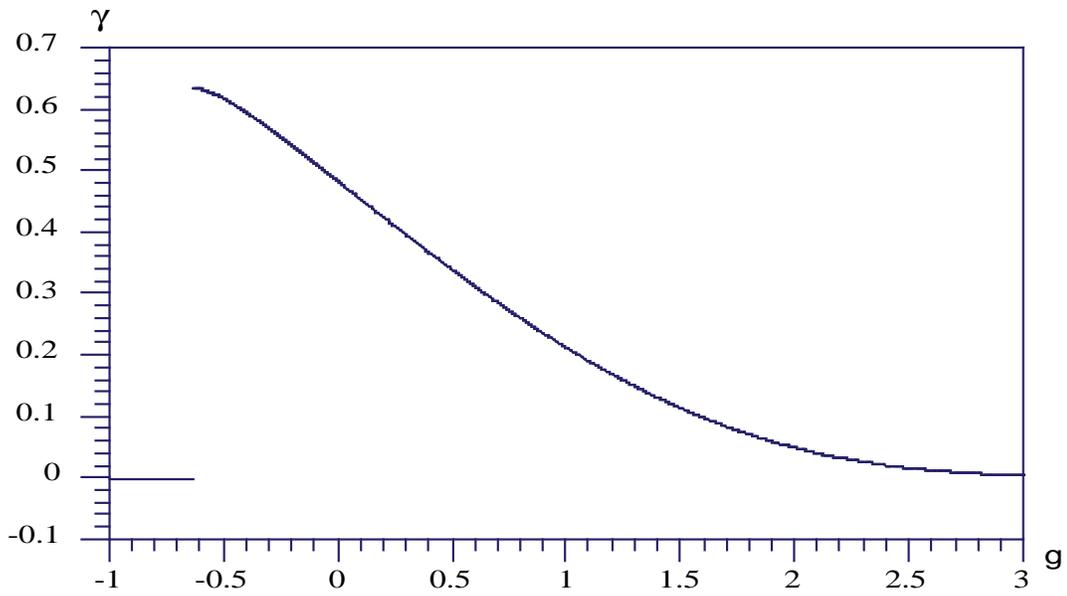}
\caption{Behaviour of $\gamma_0$
($=\lim_{T\to0}\beta\overline{(\langle\tau^2\rangle-\langle\tau\rangle^2)}$~),
in the replica symmetric solution, as a function of external field.}
\label{gam0-vs-g}
\end{figure}
We show in fig. \ref{q0-vs-g} and \ref{gam0-vs-g} $ q~(=\rho)$ and $\gamma_0$
as functions of external field. As expected both are discontinuous on the
transition point. Next we look to thermodynamical functions, the internal
energy and entropy densities are given by:
\[
 u=-(g+\gamma_0)q
\]
\beq
 s=-\frac{1}{4}\gamma_0^2 ~.
\label{entr-T0}
\eeq

The entropy (\ref{entr-T0}) is negative in the range $g>g_0$ where $\gamma_0$ is
different from zero, so we should expect the Replica Symmetry to be broken in
this region. We stress that, differently from SK case, we have a region in which
the Replica Symmetric Solution remains physical down to zero temperature. The
maximum absolute value of the zero temperature entropy is at $g=g_0^+$ where it
takes the value $0.101$ and it strongly decreases for higher values of $g$
(e.g. at $g=1 ~~ s=-0.011$).

 For $\beta\gg1$ the solution of (\ref{Qsp-sim}) can behave in two different
ways. In the range $g<g_0$, where $q_0~(\equiv q(T;g)|_{T=0})$ and
$\gamma_0(g)$ are identically zero, we find that all their temperature
derivatives vanish for $T\to0$, $q$ vanishes as $e^{\beta g}$ ($g<0$) and
$\gamma~(\equiv\beta(\rho-q))$ as $\beta e^{\beta g}$. Otherwise if $g>g_0$,
they depend linearly on $T$, indeed we get:
\[
 q=q_0-\frac{q_0\gamma_0(q_0+\unmezzo\gamma_0\alpha_0)}
 {(q_0+\unmezzo\gamma_0\alpha_0)^2+\frac{1}{4}\gamma_0(q_0-\alpha_0^2)}~T
\]
\[
 \gamma=\gamma_0+\frac{\unmezzo\gamma_0^2(q_0-\alpha_0^2)}
 {(q_0+\unmezzo\gamma_0\alpha_0)^2+\frac{1}{4}\gamma_0(q_0-\alpha_0^2)}~T
\]
~
\[
 \rho=q+\gamma T ~.
\]
~
\begin{figure}
 \epsfxsize=400pt\epsffile{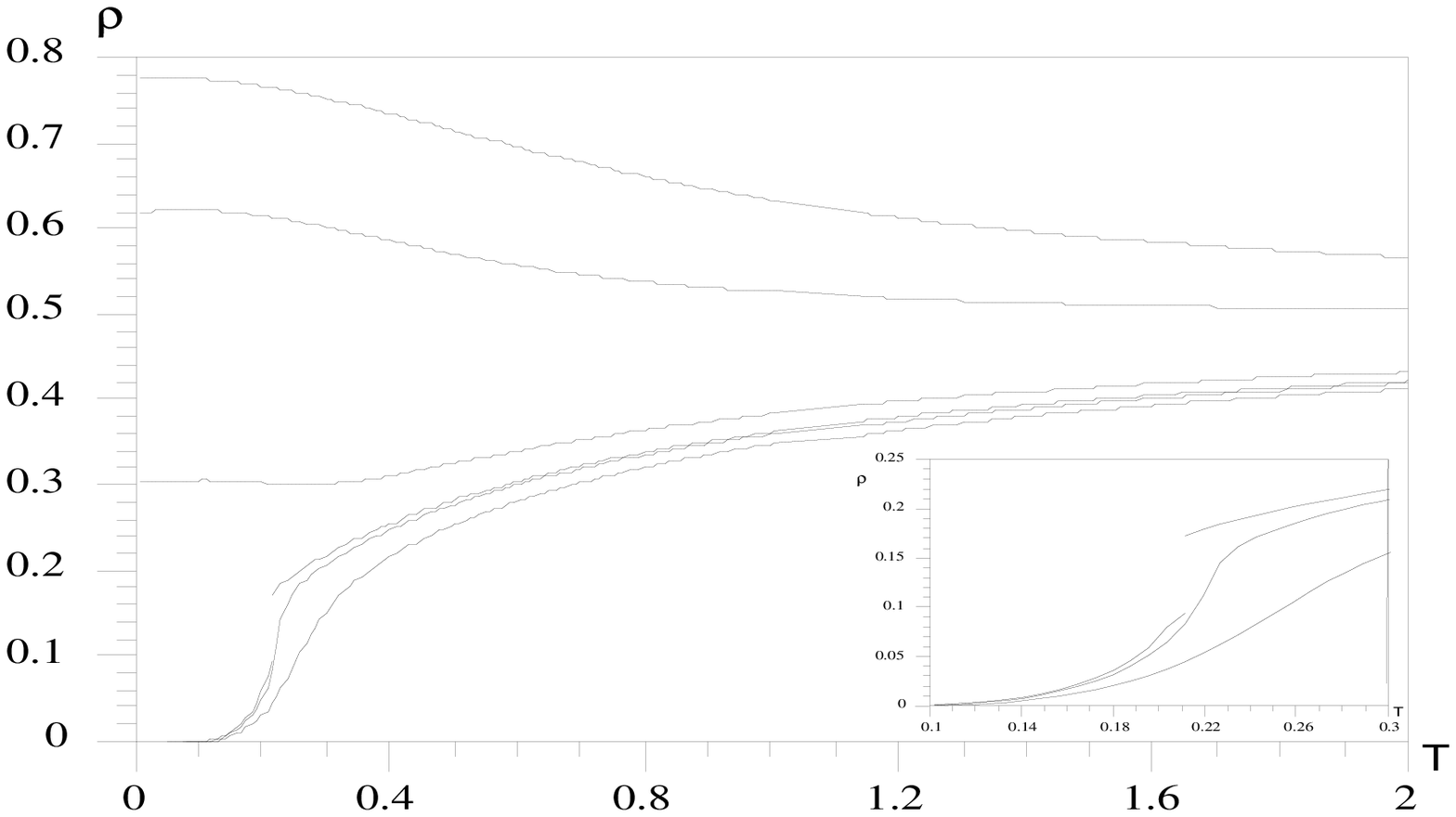}
\caption{Replica symmetric order parameter $\rho$ as function of temperature
for several values of external field. From top to bottom the values of $g$
are $0.5$, $0$, $-0.6$, $-0.68$, $-0.7$ and $-0.75$. The embedded picture
represents a magnification of the region around the transition.}
\label{rho-vs-T}
\end{figure}
\begin{figure}
 \epsfxsize=400pt\epsffile{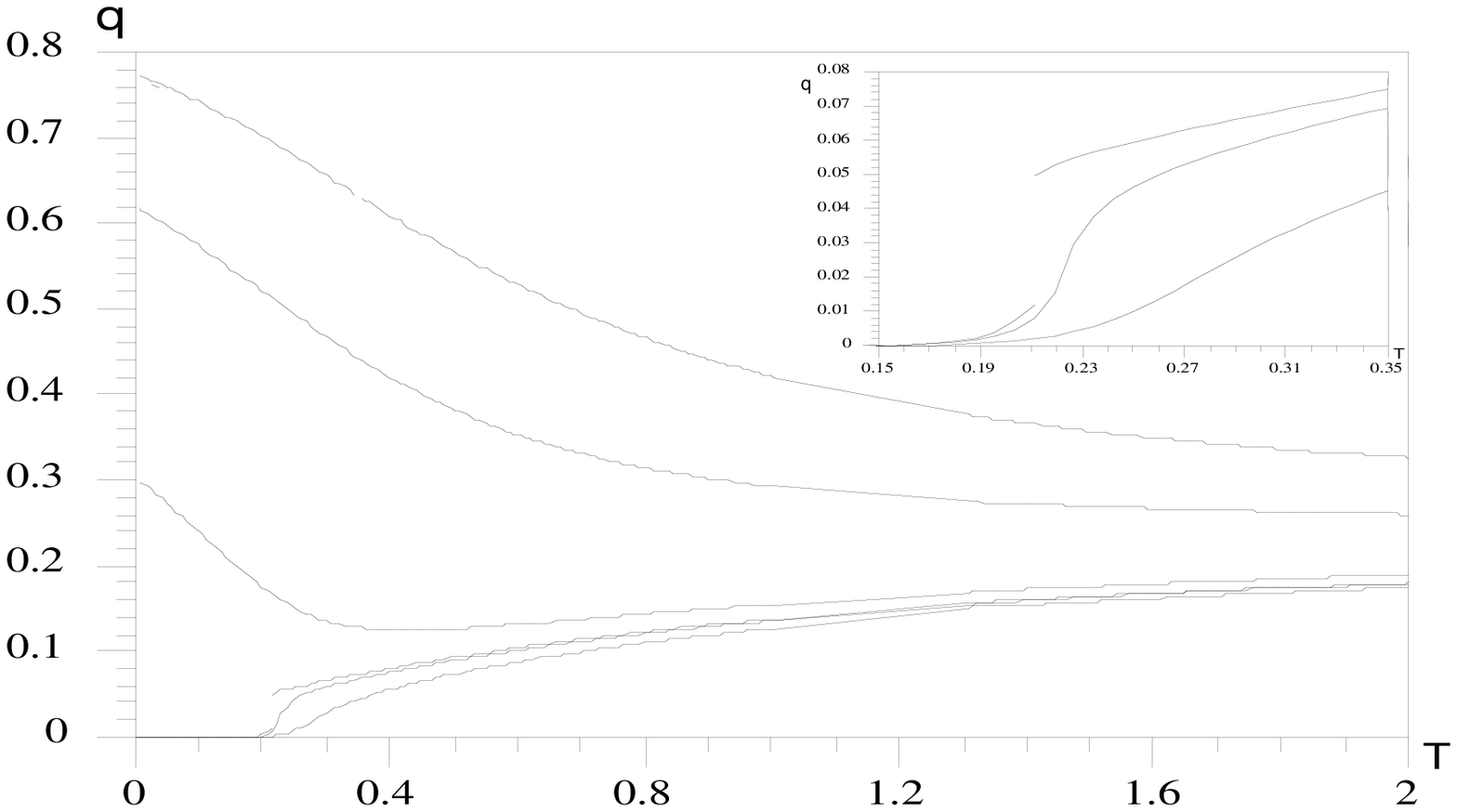}
\caption{Replica symmetric order parameter $q$ as function of temperature for
several values of external field. From top to bottom the values of $g$ are $0.
5$, $0$, $-0.6$, $-0.68$, $-0.7$ and $-0.75$. The embedded picture represents
a magnification of the region around the transition.}
\label{q-vs-T}
\end{figure}
We have also numerically solved (by iteration) equations (\ref{Qsp-sim})
for several values of $T$ and $g$, results are plotted in fig. \ref{rho-vs-T}
and \ref{q-vs-T}.
\begin{figure}
 \epsfxsize=400pt\epsffile{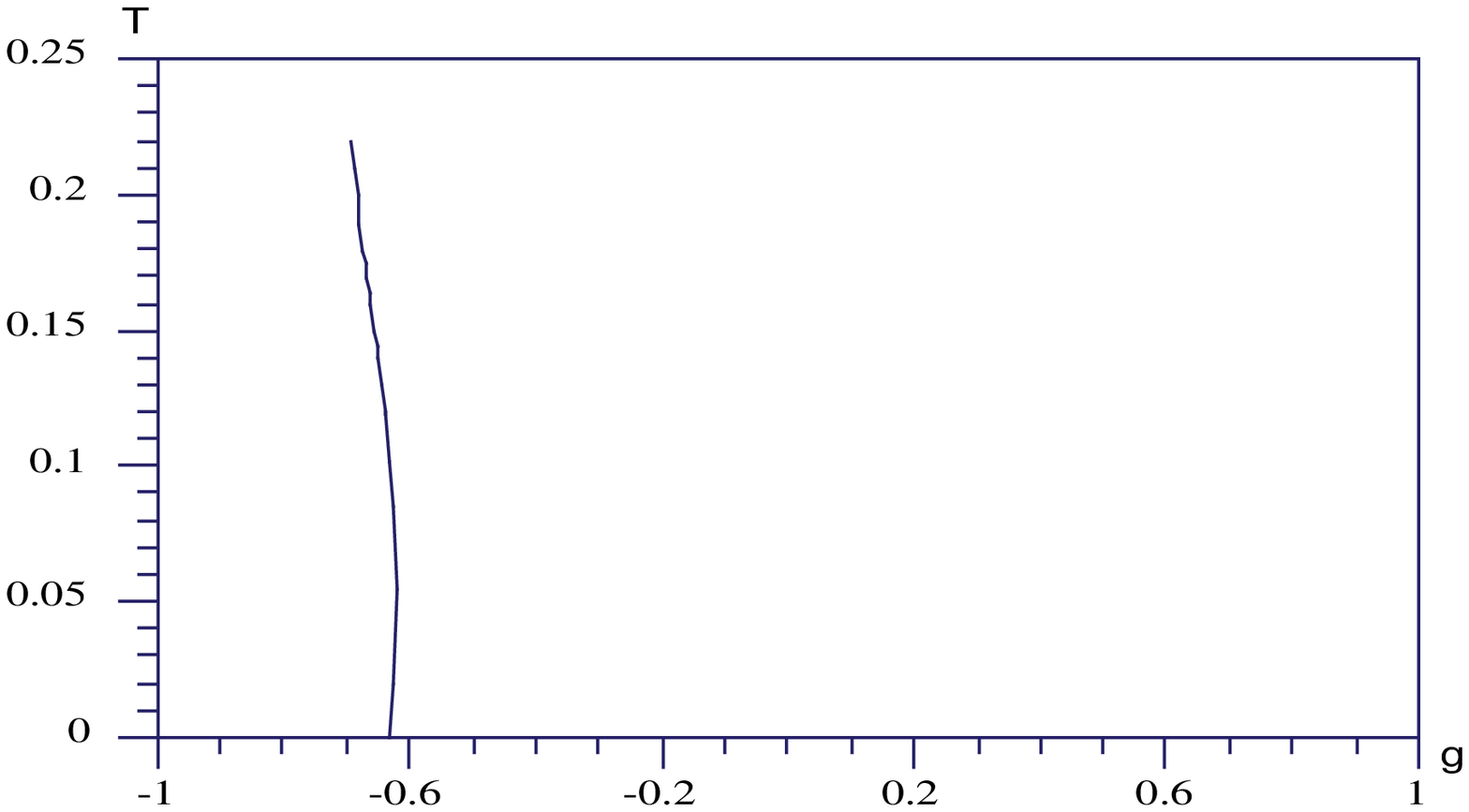}
\caption{Phase diagram for the replica symmetric solution on the plane
$(g;T)$. A line of first order phase transitions ends with a second order
transition point.}
\label{D-fase}
\end{figure}
We find a line of first-order phase transitions in the 
phase diagram of fig. \ref{D-fase}.
\begin{figure}
 \epsfxsize=400pt\epsffile{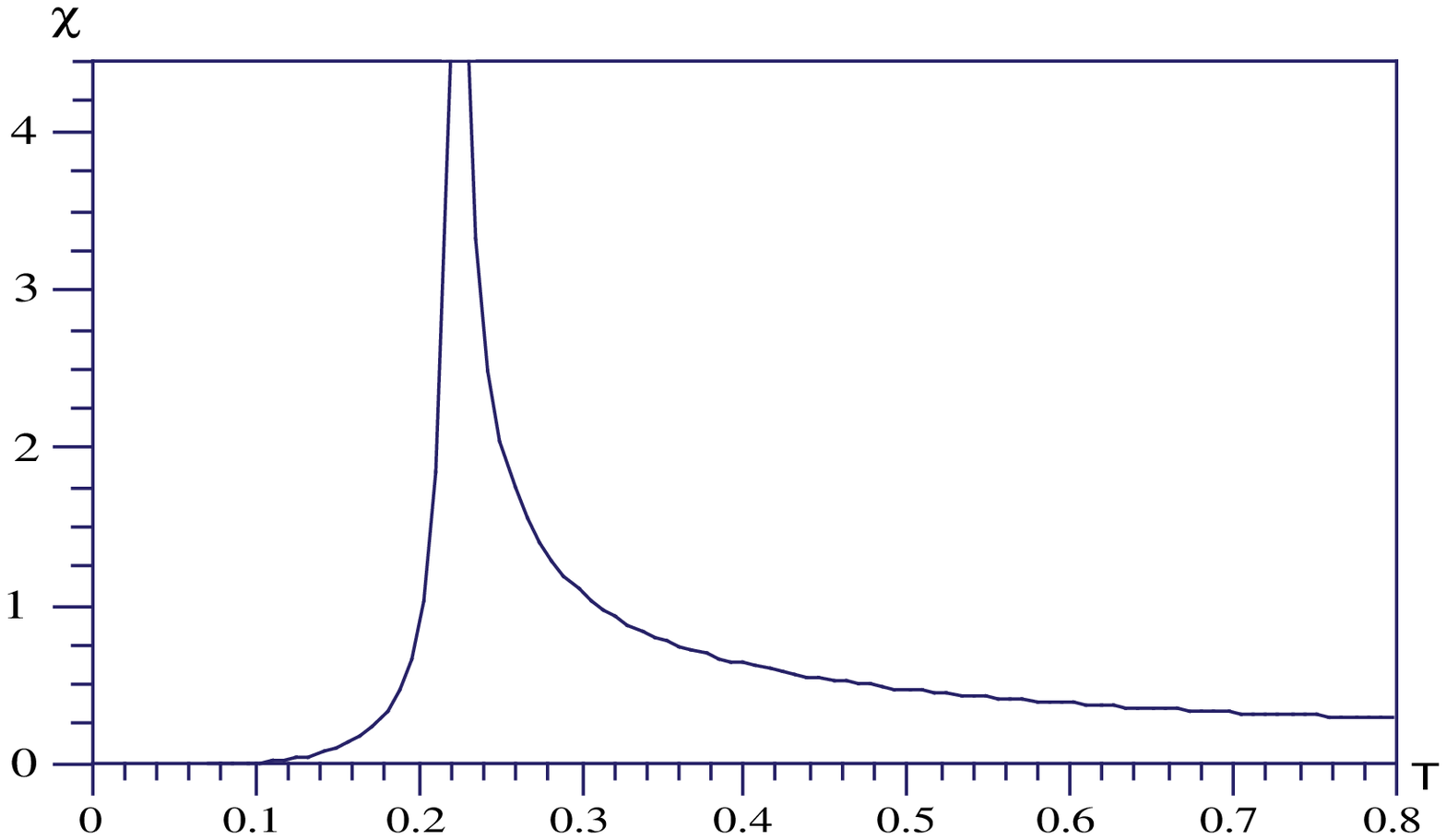}
\caption{Linear Response Function $\chi=\partial{\rho}/\partial{g}$
(``susceptibility'') as a function of temperature for $g=-0.7$.}
\label{chi-vs-T}
\end{figure}
Such line ends in a second-order transition point for $T\simeq0.22$ and
$g\simeq-0.7$, as attested from fig. \ref{chi-vs-T} where the linear response
function $\chi=\partial{\rho}/\partial{g}$ (the ``susceptibility'') is plotted
versus temperature for $g=-0.7$.

\section{Study of stability}

~ As we pointed out in the previous section the Replica Symmetric ansatz for the
saddle point equations (\ref{Qsp}) can lead to unphysical results at low
temperature. This should be considered as a signal\cite{MPV} of Replica
Symmetry Breaking (RSB). To check this conjecture we have to calculate the
eigenvalues of the Hessian of (\ref{A[Q]}), this will be done in this section
following the line of ref. \cite{DEALM} but with two main differences. The
first refers to the fact that  we have $\overline{\phi}=0$ but the diagonal
part of the matrix $Q_{ab}$ is different from zero and plays the same role as
the magnetization vector of ref. \cite{SK1,SK2,DEALM} where
$\overline{J}=J_0\neq0$. The second difference is that we never impose a priori
$Q_{ab}$ to be a symmetric matrix. As a matter of fact the only term in $A[Q]$
depending on anti-symmetric part of $Q_{ab}$ is $\tra Q^2$, its contribution
could be integrated out and has no physical significance but allowing for its
presence permits a more compact and elegant notation. We can indeed treat the
diagonal and the off-diagonal part of $Q_{ab}$ in the same way. However we
shall see that the saddle point is (as it should be) always a symmetric matrix
and stable against anti-symmetric perturbation.

The Hessian of $A[Q]$ is:
\[
 H_{ab}^{cd}=
 \frac{\partial^2 A}{\partial Q_{ab}\partial Q_{cd}}=
 \unmezzo\beta^2\delta_{ac}\delta_{bd}+\Gamma_{ab}^{cd} ~,
\]
where we have set:
\[
 \Gamma_{ab}^{cd}=\unquarto\beta^4\left[
 \langle\tau^a\tau^b\rangle_Q\langle\tau^c\tau^d\rangle_Q-
 \langle\tau^a\tau^b\tau^c\tau^d\rangle_Q\right]
\]
and the elements of $\Gamma$ have the following symmetries:
\[
 \Gamma_{ab}^{cd}=\Gamma_{ba}^{cd}=\Gamma_{ab}^{dc}=\Gamma_{cd}^{ab} ~.
\]
 Remembering that $\tau^2=\tau$, and substituting the Replica Symmetric ansatz
for the saddle point, we have the following different matrix elements for
$\Gamma$:
\[
 \Gamma_{aa}^{aa}=\unquarto\beta^4\left[
 \langle\tau^a\rangle_Q^2-\langle\tau^a\rangle_Q\right]
 =\unquarto\beta^4\left(\rho^2-\rho\right)
\]
\[
 \Gamma_{aa}^{cc}=\unquarto\beta^4\left[
 \langle\tau^a\rangle_Q\langle\tau^c\rangle_Q
 -\langle\tau^a\tau^c\rangle_Q\right]
 =\unquarto\beta^4\left(\rho^2-q\right)
\]
\[
 \Gamma_{ab}^{aa}=\unquarto\beta^4\left[
 \langle\tau^a\tau^b\rangle_Q\langle\tau^a\rangle_Q
 -\langle\tau^a\tau^b\rangle_Q\right]
 =\unquarto\beta^4\left(q\rho-q\right)
\]
\[
 \Gamma_{ab}^{cc}=\unquarto\beta^4\left[
 \langle\tau^a\tau^b\rangle_Q\langle\tau^c\rangle_Q
 -\langle\tau^a\tau^b\tau^c\rangle_Q\right]
 =\unquarto\beta^4\left(q\rho-\rho_3\right)
\]
\[
 \Gamma_{ab}^{ab}=\unquarto\beta^4\left[
 \langle\tau^a\tau^b\rangle_Q^2
 -\langle\tau^a\tau^b\rangle_Q\right]
 =\unquarto\beta^4\left(q^2-q\right)
\]
\[
 \Gamma_{ab}^{ad}=\unquarto\beta^4\left[
 \langle\tau^a\tau^b\rangle_Q\langle\tau^a\tau^d\rangle_Q
 -\langle\tau^a\tau^b\tau^d\rangle_Q\right]
 =\unquarto\beta^4\left(q^2-\rho_3\right)
\]
\[
 \Gamma_{ab}^{cd}=\unquarto\beta^4\left[
 \langle\tau^a\tau^b\rangle_Q\langle\tau^c\tau^d\rangle_Q
 -\langle\tau^a\tau^b\tau^c\tau^d\rangle_Q\right]
 =\unquarto\beta^4\left(q^2-\rho_4\right) ~,
\]
where explicitly written replica labels are different. In the $n\to0$
limit, the required expectation values are:
\[
 \langle\tau^a\rangle_{Q^*}=Q_{aa}^*=\rho
\]
\[
 \langle\tau^a\tau^b\rangle_{Q^*}=Q_{ab}^*=q
\]
\[
 \langle\tau^a\tau^b\tau^c\rangle_{Q^*}\equiv\rho_3=
 (2\pi)^{-\unmezzo}\int\left[1+e^{-\beta(\alpha+z\sqrt{q})}\right]^{-3}
 e^{-\unmezzo z^2}~dz
\]
\nopagebreak
\[
 \langle\tau^a\tau^b\tau^c\tau^d\rangle_{Q^*}\equiv\rho_4=
 (2\pi)^{-\unmezzo}\int\left[1+e^{-\beta(\alpha+z\sqrt{q})}\right]^{-4}
 e^{-\unmezzo z^2}~dz~.
\]
The eigenvalues equation for the Hessian $H^{cd}_{ab}$:
\beq
 \sum_{c,d}H^{cd}_{ab}\eta_{cd}=\lambda\eta_{ab}~,
\label{eq-aut}
\eeq
has, for general $n$, four classes of eigenvectors with no more than six
distinct eigenvalues. Anti-symmetric eigenvectors ($\eta_{ba}=-\eta_{ab}$) give
the eigenvalue $\lambda_0=\unmezzo\beta^2$ which (as we should expect) is always
positive. Eigenvectors that are symmetric matrices, and invariant under the
Replica Group, give two eigenvalues that, in the $n\to0$ limit, are:
\beq
 \lambda_{1,2}=\unmezzo\beta^2+
 \frac{1}{8}\beta^4(8\rho_3-6\rho_4-q-\rho)\pm\frac{1}{8}\beta^4
 \sqrt{(8\rho_3-6\rho_4-3q+\rho)^2-16(\rho_3-q)^2}.
\label{aut-inv}
\eeq\\
Symmetric eigenvectors ($\eta_{ba}=\eta_{ab}$) that are invariant under
interchange of all but one of the replicas give, for general $n$, two more
eigenvalues $\lambda_{3,4}$ that for $n\to0$ reduce to the previous ones. There
are finally symmetric eigenvectors that are invariant under interchange of all
but two replicas, for $n\to0$ these give rise to the eigenvalue:
\beq
 \lambda_5=\unmezzo\beta^2\left[1+\beta^2(2\rho_3-\rho_4-q)\right].
\label{lambda5}
\eeq
~ In the high temperature regime the eigenvalues $\lambda_1$ and $\lambda_2$ are
found to be, in spite of the hermiticity of the Hessian matrix, complex
conjugate. This should not be too surprising because we are working in a
space where the norm is not positive definite ($\lim_{n\to0}\frac{1}{n}\tra
Q^2=\rho^2-q^2$). However, as could be seen considering the gaussian
approximation to the integral in (\ref{zeta-n3}), we believe that the stability
of the solution is determined by their (common) real part which is always
positive. At lower temperature these eigenvalues are real and never negative,
although they are found to vanish linearly on the second-order transition
point. This means that this saddle-point is stable against
{\em Replica-Symmetric\/} perturbation.

\begin{figure}
 \epsfxsize=400pt\epsffile{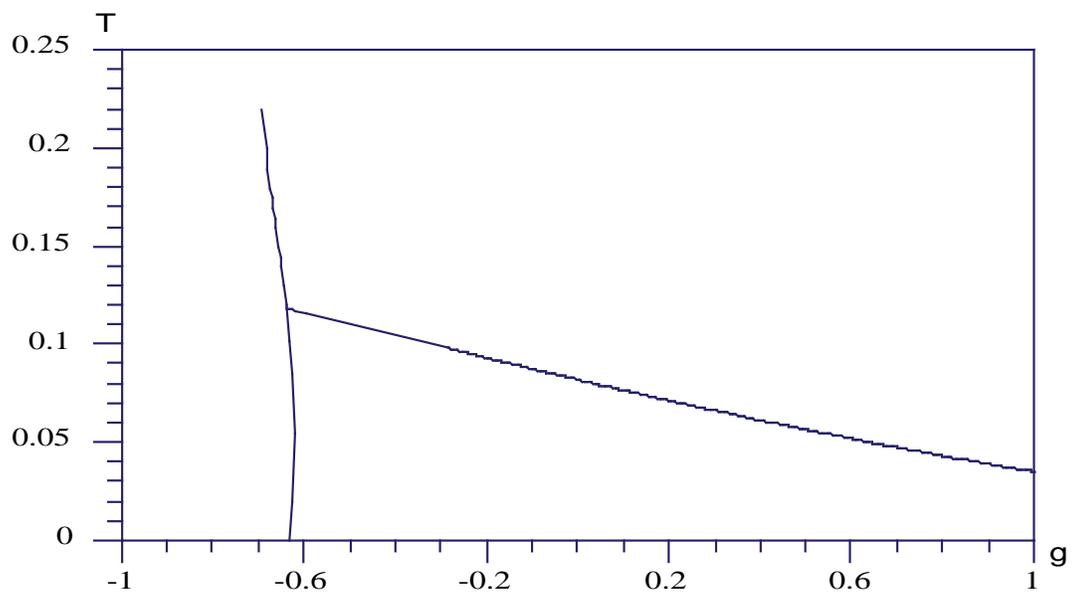}
\caption{Limits of stability for the replica symmetric solution superimposed on
its phase diagram. The region under the instability line is the Glassy Phase.}
\label{DATh-mio}
\end{figure}

 The eigenvalues $\lambda_0$ $\lambda_3$ and $\lambda_4$ are not relevant and
thus the stability of this solution against Replica-Symmetry-Breaking
perturbations is determined by the eigenvalue $\lambda_5$. This eigenvalue is
always real, to study its sign it is useful to define $\Lambda_5=2T^3\lambda_5$,
so we get:
\beq
 \Lambda_5=T-\unottavo(2\pi q)^{-\unmezzo}
 \int\frac{e^{-\frac{(\alpha+2Tu)^2}{2q}}}{\cosh^4 u}du ~.
\label{L5}
\eeq
The low temperature limit then follows in a straightforward way and we find:
\[
 \lim_{T\to0}\Lambda_5=-\frac{1}{6}(2\pi q)^{-\unmezzo}
 e^{-\frac{\alpha^2}{2q}}=-\frac{1}{6}\gamma_0~.
\]
It is therefore apparent that, if $g\geq g_0$, $\lambda_5$ must become
negative at low enough temperatures. The {\em Replica-Symmetric\/} saddle
point is unstable and the Replica Symmetry is {\em spontaneously broken\/}.
\begin{figure}
 \epsfxsize=400pt\epsffile{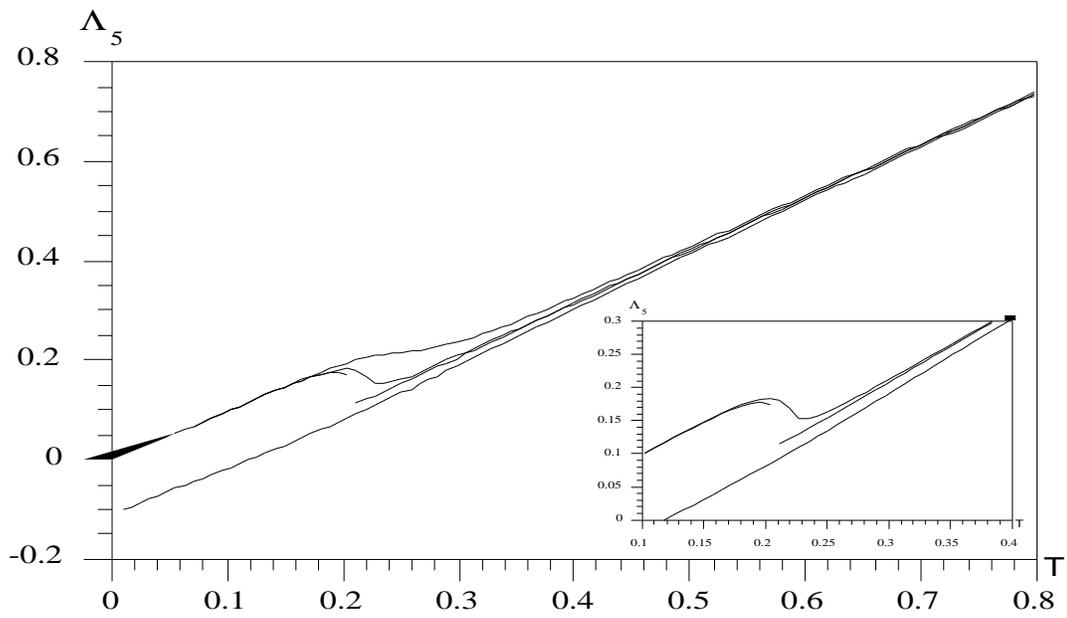}
\caption{Behaviour of $\Lambda_5$ ($=2T^3\lambda_5$) as a function of
temperature for several values of external field. From bottom to top  the
values of $g$ are $-0.6$, $-0.68$, $-0.7$ and $-0.75$. The embedded picture
represents a magnification of the region around the transition.}
\label{L5-vs-T}
\end{figure}
We plot in figure \ref{L5-vs-T} the behaviour of  $\Lambda_5$ versus
temperature for several values of $g$. The line of instability obtained
by numerical evaluation is shown in figure \ref{DATh-mio}, we stress the
presence of a region in which the Replica Symmetry remains {\em exact\/}
down to zero temperature.

\section{Zero temperature results with first order Replica Symmetry Breaking}

~ Once stated that the Replica Symmetry can be broken, we looked for saddle
points that are not invariant under the Replica Group. As a first step we
analyzed the zero temperature behaviour of the solution given by the ansatz 
proposed in ref. \cite{par79,par80-1}. Let us divide the $n$ replicas in
$\nu\equiv n/m$ groups, each of $m$ replicas, and take the matrix elements
of $Q$ as follows:
\[
 Q_{ab}=\rho=q+d+b ~~~~~~~ \mbox{if}~~~ a=b
\]
\[
 Q_{ab}=q_1=q+d ~~~~~~~\mbox{if}~~~ a\neq b ~~~\mbox{but}~~ I(a/m)=I(b/m)
\]
\beq
 Q_{ab}=q_0\equiv q ~~~~~~\mbox{if}~~~ I(a/m)\neq I(b/m) ~,
\label{ans-RSB1}
\eeq
where as usual\cite{par79} $~I(x)=\min\{n\in N:n\geq x\}$.
With this position we have $\tra Q^2=n\left[\rho^2+(m-1)q_1^2+(n-m)q_0^2\right]$
and the effective Hamiltonian (\ref{HQ[tau]}) reads:
\beq
 H_Q[\tau]=-\beta\alpha\sum_a\tau^a-
 \unmezzo\beta^2 q_0\left(\sum_a\tau^a\right)^2-
 \unmezzo\beta^2 d\sum_{l} \left(\sum_{k(a)=l}\tau^a\right)^2
\label{HQ-RSB1}
\eeq
where $k(a)=I(a/m)$ and as previously we have set $\alpha=g+\unmezzo\beta b$.

 Using standard properties of Gaussian integrals, and setting
$g_\sigma(x)=(2\pi\sigma)^{-\unmezzo}e^{-\frac{x^2}{2\sigma}}$,
we find in the $n\to0$ limit:
\beq
 f=\unquarto\beta\left[\rho^2+(m-1)q_1^2-mq_0^2\right]-
 \frac{1}{\beta m}\int g_{q_0}(z)\ln\left[I_m(\alpha+z;d)\right]dz ~,
\label{f(q)-RSB1}
\eeq
where:
\beq
 I_m(x;d)=\int\left[1+e^{\beta(x+y)}\right]^m g_d(y)dy ~.
\label{Im(z;d)}
\eeq
Substituting the ansatz (\ref{ans-RSB1}) in the saddle point equations
(\ref{Qsp}) we find, in the $n\to0$ limit, the following equations for
$\rho$, $q_1$  and $q_0$:
\[
 \rho=\int g_{q_0}(z-\alpha)\frac
 {\int\left[1+e^{\beta(z+y)}\right]^{m-1}e^{\beta(z+y)}g_d(y)dy}
 {I_m(z;d)}dz
\]
\beq
 q_1=\int g_{q_0}(z-\alpha)\frac
 {\int\left[1+e^{\beta(z+y)}\right]^{m-2}e^{2\beta(z+y)}g_d(y)dy}
 {I_m(z;d)}dz
\label{Qsp-RSB1}
\eeq
\[
 q_0=\int g_{q_0}(z-\alpha)\left[\frac
 {\int\left[1+e^{\beta(z+y)}\right]^{m-1}e^{\beta(z+y)}g_d(y)dy}
 {I_m(z;d)}\right]^2 dz ~.
\]

To decide if the correct saddle point has to be a maximum or a minimum of eq.
(\ref{f(q)-RSB1}) we note that 
$\lim_{n\to0}\frac{1}{n}~\tra Q^2=\rho^2-(1-m)q_1^2-mq_0^2$ so, for
$0\leq m\leq1$, we should expect the saddle point to be a minimum with respect
to $\rho$ (diagonal parameter) and a maximum with respect to $q_1$ and $q_0$
(off-diagonal parameters). Because $m$ itself is a parameter for the
off-diagonal part of $Q_{ab}$ we speculate that $f$ should be maximized with
respect to it\cite{MPV}.

\begin{figure}
 \epsfxsize=400pt\epsffile{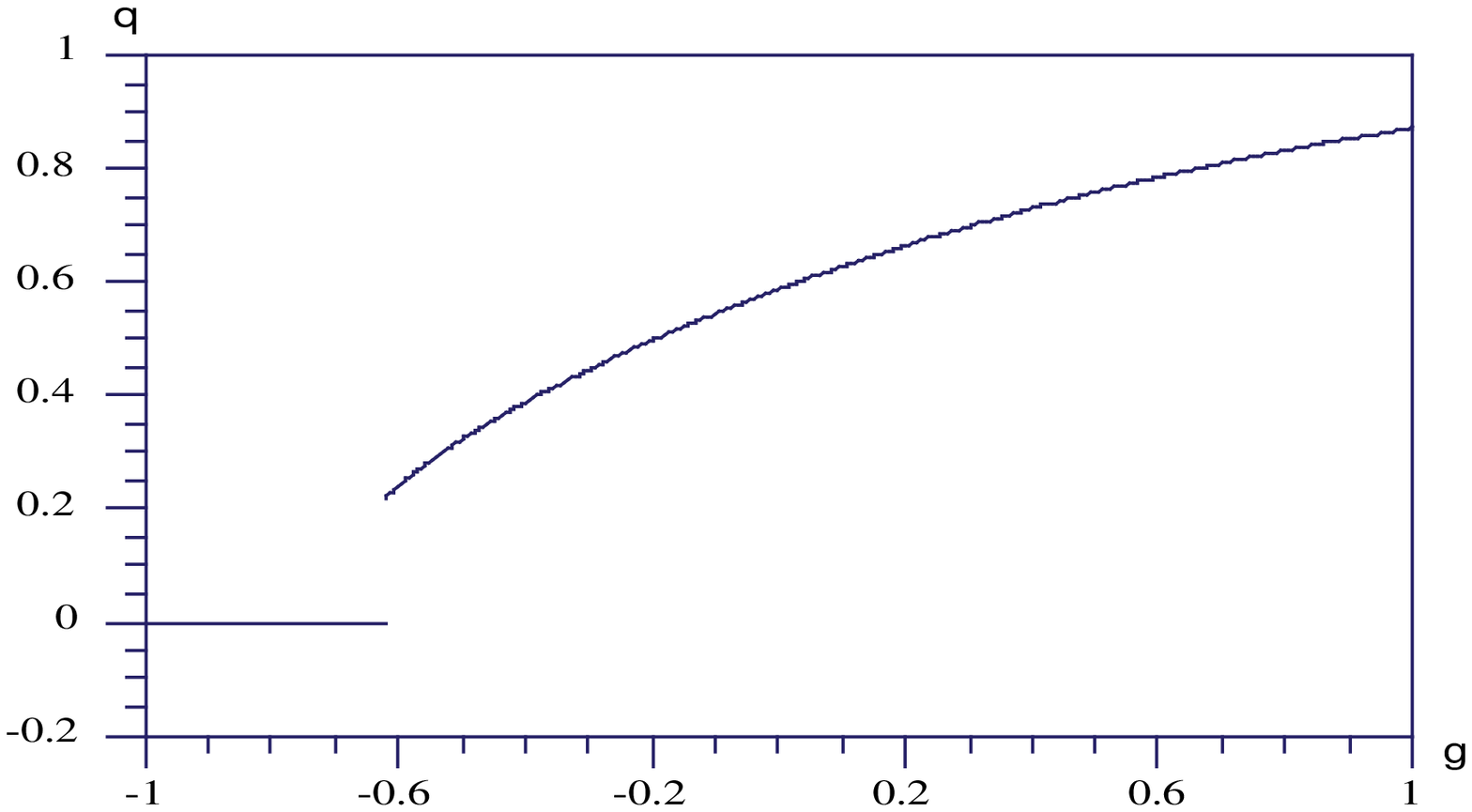}
\caption{Solution with first order replica symmetry breaking:
Off-diagonal order parameter $q$ ($=q_0$) at $T=0$ versus external field.}
\label{q0-vs-g-RSB1}
\end{figure}

\begin{figure}
 \epsfxsize=400pt\epsffile{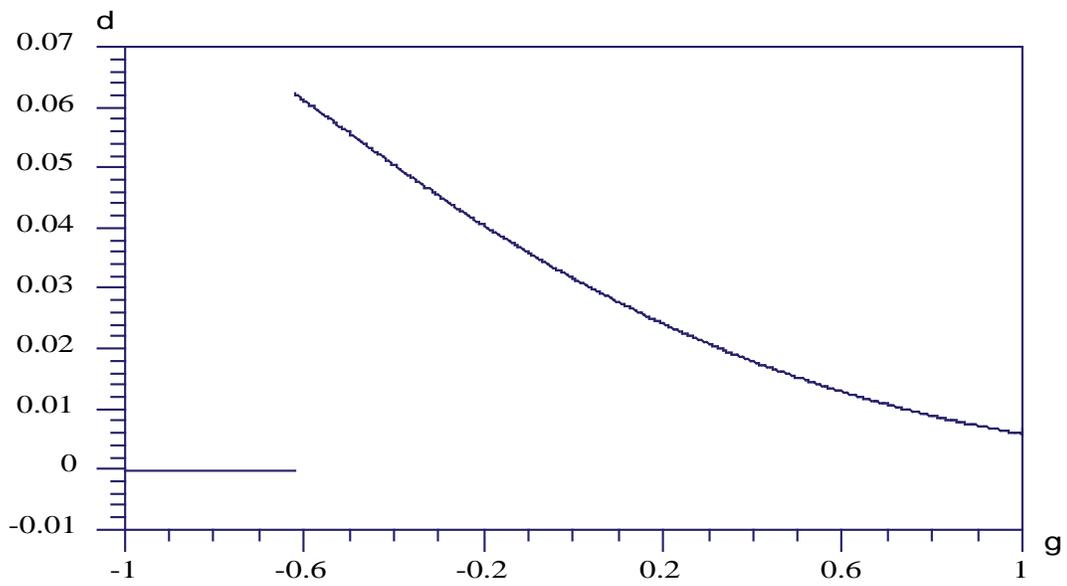}
\caption{Solution with first order replica symmetry breaking:
Symmetry breaking parameter $d$ ($=q_1-q_0$) as a function of external field
at zero temperature.}
\label{d0-vs-g}
\end{figure}

\begin{figure}
 \epsfxsize=400pt\epsffile{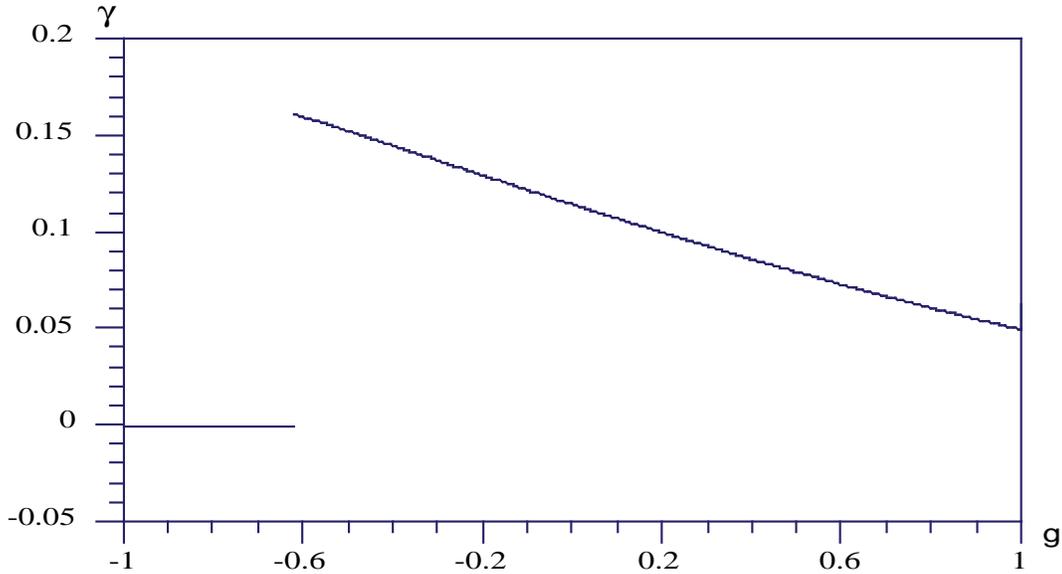}
\caption{Solution with first order replica symmetry breaking:
Behaviour of $\gamma$ ($=\lim_{T\to0}\beta(\rho-q_1)$) as a function of
external field.}
\label{gam0-vs-g-RSB1}
\end{figure}

In order to work out the $T\to0$ limit of eq. (\ref{Qsp-RSB1}) let us consider
the internal energy density:
\[
 u=-g\rho-\unmezzo\beta b(\rho+q_1)-\unmezzo\beta md(q_0+q_1)~.
\]
To keep $u$ finite as $\beta\to\infty$ we set $b=\gamma T$ and $m=\mu T$, and we
expect that $\gamma$ and $\mu$ remain finite as $T\to0$, indeed in this limit
eq. (\ref{Qsp-RSB1}) leads to:
\[
 \gamma=\int\frac{g_d(z)g_{q_0}(\alpha+z)}
 {\erf\left(\frac{z}{\sqrt{d}}\right)+
 e^{-\mu(z-\unmezzo\mu d)}\erf\left(\frac{\mu d-z}{\sqrt{d}}\right)}dz
\]
\nopagebreak
\beq
 \rho\equiv q+d=\int\frac
 {e^{-\mu(z-\unmezzo\mu d)}\erf\left(\frac{\mu d-z}{\sqrt{d}}\right)}
 {\erf\left(\frac{z}{\sqrt{d}}\right)
 +e^{-\mu(z-\unmezzo\mu d)}\erf\left(\frac{\mu d-z}{\sqrt{d}}\right)}
 g_{q_0}(\alpha+z)dz
\label{Qsp-RSB1-T0}
\eeq
\nopagebreak
\[
 q=\int\left[\frac
 {e^{-\mu(z-\unmezzo\mu d)}\erf\left(\frac{\mu d-z}{\sqrt{d}}\right)}
 {\erf\left(\frac{z}{\sqrt{d}}\right)
 +e^{-\mu(z-\unmezzo\mu d)}\erf\left(\frac{\mu d-z}{\sqrt{d}}\right)}
 \right]^2 g_{q_0}(\alpha+z)dz ~.
\]
In the same limit the free energy density (\ref{f(q)-RSB1}) then becomes:
\beq
 f=\frac{1}{4}\left[2\gamma(q+d)+\mu d(2q+d)\right]-
 \frac{1}{\mu}\int g_{q}(\alpha+z)\ln\left[I_m(z;d)\right]dz ~,
\label{free-RSB1-T0}
\eeq
where:
\beq
 \lim_{T\to0}I_m(z;d)={\erf\left(\frac{z}{\sqrt{d}}\right)
 +e^{-\mu(z-\unmezzo\mu d)}\erf\left(\frac{\mu d-z}{\sqrt{d}}\right)}~.
\label{Im-T0}
\eeq
~ Equations (\ref{Qsp-RSB1-T0}) have been numerically solved by iteration with
$\mu$ held fixed, next the resulting free energy (\ref{free-RSB1-T0}) has
been maximized with respect to $\mu$ using a standard IMSL routine.
\begin{figure}
 \epsfxsize=400pt\epsffile{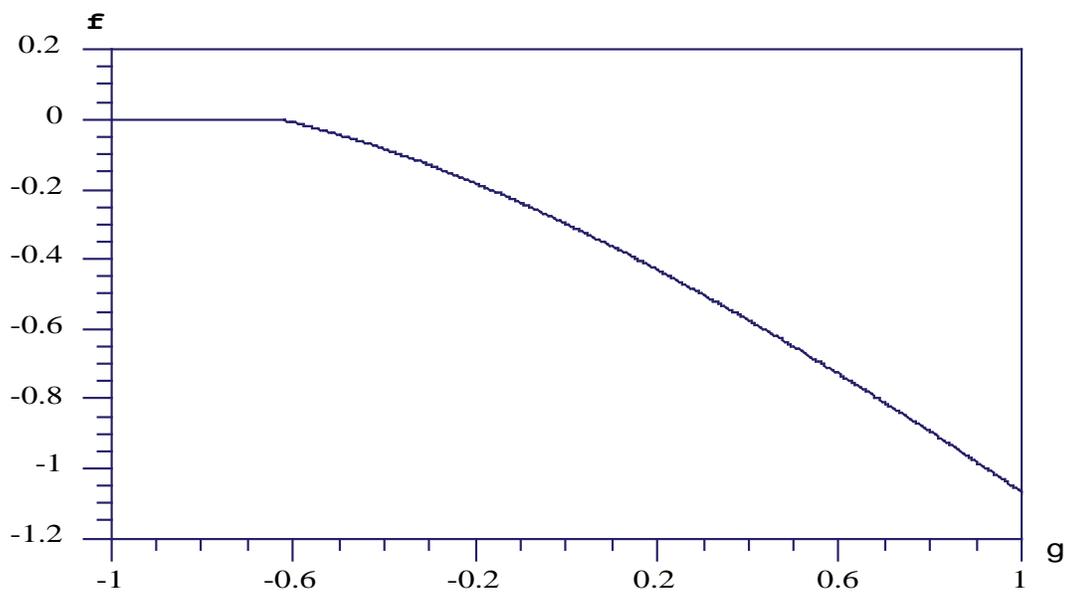}
\caption{Free energy density versus external field at zero temperature with
first order replica symmetry breaking. The first order transition point
is at $g=-0.625$.} 
\label{f0-vs-g-RSB1}
\end{figure}
\begin{figure}
 \epsfxsize=400pt\epsffile{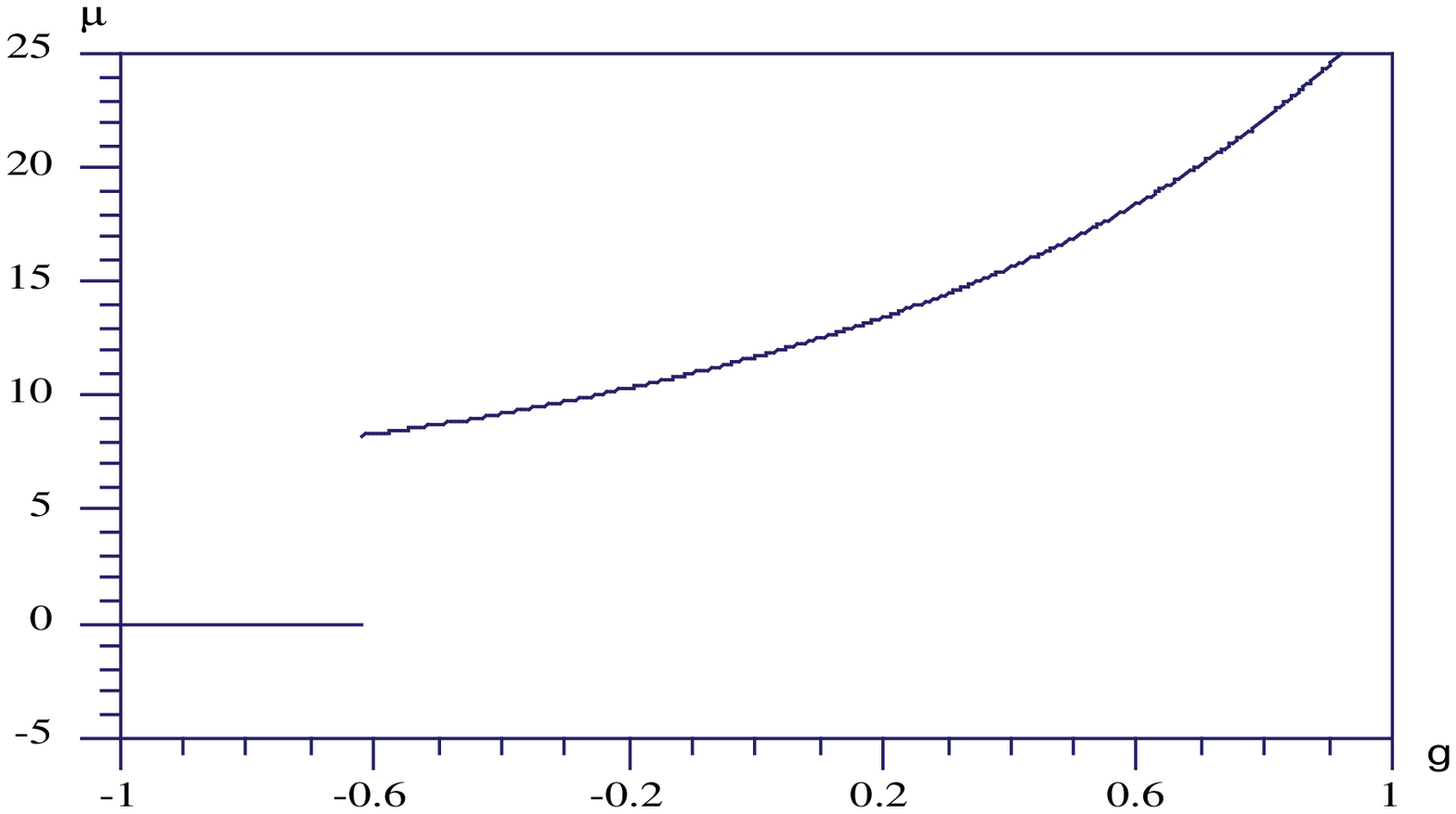}
\caption{Behaviour of $\mu$ ($=\lim_{T\to0}\beta m$) as a function
of external field.} 
\label{mu0-vs-g}
\end{figure}
Such numerical solution of the saddle point problem gives the results plotted
in figures \ref{f0-vs-g-RSB1}-\ref{gam0-vs-g-RSB1} as functions of $g$.
We note that the transition point is shifted from $g_0=-0.63633$ to
$g_0\simeq-0.6250$  and the discontinuity of $q$ ($0.27$) is split in
$0.28$ for $\rho$ and $0.22$ for $q_0$; we also note that the values
of $\gamma$ are about a factor four smaller than those obtained in the
Replica Symmetric approximation.

\section{Discussion conclusions and outlook}

~ In this paper we have investigated the thermodynamical properties of a 
lattice-gas model with infinite-ranged random interactions. We have discussed
its relation with SK model and showed that the {\em averaged thermodynamical
functions\/} of the two systems are not directly related to each other.

In SK case the zero-field Hamiltonian has a {\em global $Z_2$ symmetry\/} and
all thermodynamical functions are either even or odd in the external field. SK
model can have a first-order phase transition, as a function of magnetic field,
only if a strong enough ferromagnetic part (i.e. a non-zero mean) is added to
the random coupling, such a transition is indeed related to the breaking of the
global $Z_2$ symmetry induced (as in the homogeneous case) by a ferromagnetic
coupling. Moreover, at each value of external field, SK's Replica Symmetric
Solution always becomes unstable for low enough temperature.

It is apparent how our picture is different from the usual one. We have
considered the case of purely (zero-mean) random interactions and our
Hamiltonian has no $Z_2$ symmetry. Nevertheless the {\em Replica Symmetric
Solution\/} of our system exhibits a line of first-order phase transition
points ending with a second-order transition point. This feature is robust to
{\em Replica Symmetry Breaking\/}, indeed the second-order transition point
lies outside the unstable region and we only expect that the transition line
should be deflected on the instability boundary. Two phases co-exist along the
transition line and, as we have stressed, in one of them the Replica Symmetry
is {\em exact down to zero temperature.\/} We think that this is a consequence
of the fact that in this phase the system should behave like an homogeneous
one.

To have a physical interpretation of the obtained phase diagram it is useful
to return to a more conventional framework for lattice gases. The effective
Hamiltonian (\ref{Ham}) describes a system of particles, in a discrete space,
interacting with a two body potential $\phi_{kl}$. In this context $g$ is
closely related to the chemical potential and the thermodynamic potential
(\ref{f-med}) is the negative of the pressure. The parameters $\rho$ and $q$
are respectively the quenched averages of particle density and of its square.
When a first-order transition occurs there are two coexisting phases, the
low-density one is to be interpreted as the gas phase, while the other is the
liquid one. In our model the gas phase is that which is replica-stable at all
temperatures. On the other hand, at sufficiently low temperatures, the Replica
Symmetry breaks inside the liquid phase giving up to a glassy state.
A part of the line of first-order transitions may therefore be interpreted
as glass in equilibrium with its vapour.

We thus have a simple and soluble mean-field model accounting (in a conventional
and so far well understood way) for both a liquid-gas transition and a glassy
regime. This could be regarded as an intriguing paradigm for capturing the
structural glass transition in the framework of replica theory.

There is a flaw in this. We have analyzed a model with quenched (random)
disorder included {\em by hand\/}. Structural glasses instead do not
necessarily have random interactions in their hamiltonians. In several recent
publications (see e.g. \cite{BOUMEZ,MAPARI1,MAPARI2}) the main effort was
first to present a system, {\em without random interactions\/}, which behave
in a glassy way, and next try to construct a disordered model which mimics
the starting one. ~ Here we have to go in the opposite direction. We have a
good candidate of random system, the way of proposing a corresponding
deterministic one will be presented elsewhere\cite{RUSSO2}.

\section*{Acknowledgments}

 ~~I am very grateful to Prof. Giorgio Parisi for the help he gave in this work
with his suggestions and experience, and also for his careful and critical
reading of the original manuscript.

I acknowledge Prof. Enzo Marinari for stimulating discussions and
encouragement and also for his illuminating lectures; I thank Dr. Felix Ritort
for useful discussions on various topics related to the subject of this work.

I wish acknowledge INFN, ``Sezione di Roma {\em Tor Vergata\/}'',
for support received during this work.

%

\vfill
\end{document}